\documentclass[conference]{IEEEtran}

\usepackage{amsmath}
\usepackage{amssymb}
\usepackage{easy-todo}
\usepackage{graphicx}
\usepackage{balance}
\usepackage{tikz}
\usepackage{nicefrac}
\usepackage{hyperref}
\usepackage{overpic}
\usetikzlibrary{arrows.meta}

\newtheorem{theorem}{Theorem}
\newtheorem{corollary}{Corollary}

\newcommand{\ceilfrac}[2]{\left\lceil \frac{#1}{#2} \right\rceil}

\newcommand{\floorfrac}[2]{\left\lfloor \frac{#1}{#2} \right\rfloor}

\newcommand{\bk}{\mathbf{k}}
\newcommand{\bh}{\mathbf{h}}
\newcommand{\bC}{\mathbf{C}}
\newcommand{\bU}{\mathbf{U}}

\newcommand{\linear}[3]{\draw (0,{#3/#1}) -- ({#3/#2},0);}

\newcommand{\linOR}[5]{\linear{#2}{#3}{#1}
  \linear{#5}{#3}{#4} \draw[very thick] (0,{#4/#5}) --
  ({(#1-#2*(#1-#4)/(#2-#5))/#3},{(#1-#4)/(#2-#5)}) --
  ({#1/#3},0);}

\begin{document}

\title{Optimizing over FP/EDF Execution Times:\\
Known Results and Open Problems}

\author{Enrico Bini, University of Turin, Italy}

\maketitle

\begin{abstract}
  In many use cases the execution time of tasks is unknown and can be
  chosen by the designer to increase or decrease the application
  features depending on the availability of processing capacity. If
  the application has real-time constraints, such as deadlines, then
  the necessary and sufficient schedulability test must allow the
  execution times to be left unspecified. By doing so, the designer
  can then perform optimization of the execution times by picking the
  schedulable values that minimize any given cost.

  In this paper, we review existing results on the formulation of both
  the Fixed Priority and Earliest Deadline First exact
  schedulability constraints. The reviewed formulations are expressed
  by a combination of linear constraints, which enables then
  optimization routines.
\end{abstract}

\section{Introduction}
\label{sec:intro}

The necessity to trade the accuracy of applications with the available
processing capacity is common in many application domains. Some
notable examples are the MPEG decoding which may be made at different
degrees of detail, or the solution of optimization problems which may
be solved with different proximity to the optimal solution. Notably,
the workload of inference in neural network also belongs to this class
as one may choose the desired accuracy of the answer given by a model.

If the application has real-time constrains, then schedulability must
be taken into account. Schedulability tests, however, may not lend
themselves to this type of problem as they normally need to know the
execution times and they provide a ``yes/no'' answer to the schedulability
questions, which is unfit for optimization. Instead, schedulability
conditions which are formulated as combination of algebraic
constraints between the parameters are better suited to contexts in
which some of the parameters are unknown and are free to choose by the
designer.

In this paper, we review some existing results  on this form of
constraints when the execution times are the free variables. We
address both Fixed Priority (FP) and Earliest Deadline First (EDF)
schedulability conditions and we present some open problems in this
context.

\section{System Model}
\label{sec:model}

We consider a set $\mathcal{T} = \{\tau_1,\ldots,\tau_n\}$ of $n$
periodic \emph{tasks}. Each task $\tau_i$ is
characterized by
\begin{itemize}
\item a worst-case execution time $C_i$ (which may be called
  \emph{execution time} for simplicity),
\item a \emph{period} $T_i$, and 
\item a relative \emph{deadline} $D_i$ not greater than $T_i$
  (\emph{constrained deadline} model).
\end{itemize}
All task parameters are assumed real-valued.

In the single processor context, which is addressed in this paper,
this same model also captures tasks with sporadic releases. In such a
case, $T_i$ denotes the minimum interarrival time between two
consecutive jobs.

Each task releases an infinite sequence of \emph{jobs}. Jobs are
indexed in $\mathbb{N}$ by the order of release. We assume
$0\in\mathbb{N}$.  The \emph{release time} of the $j$-th job of
$\tau_i$ is denoted by $r_{i,j}$ and releases of consecutive jobs are
constrained by
\begin{equation}
  \label{eq:min_inter}
  \forall j\in\mathbb{N},\quad r_{i,j+1} \geq r_{i,j}+T_i.
\end{equation}
Each job must also complete not later that its \emph{absolute
  deadline} $d_{i,j}$, which is set by
\begin{equation}
  \label{eq:abs_deadline}
  d_{i,j}=r_{i,j}+D_i.
\end{equation}
Finally, we use $U_i = \frac{C_i}{T_i}$ to denote $\tau_i$'s
\emph{utilization} as indeed, it is the fraction of CPU time utilized
by $\tau_i$.

For a more compact notation, we may be using
\begin{itemize}
\item $\bC=[C_1,\ldots,C_n]$ to denote the vector of all execution
  times, and
\item $\bU=[U_1,\ldots,U_n]$ to denote the vector of utilizations.
\end{itemize}

We address single processor preemptive scheduling. This means that the
scheduler may decide to preempt a running job to schedule another
higher priority job. The interrupted job will then be continued later.

This paper considers only Fixed Priority (FP) and Earliest Deadline
First (EDF) schedulers, in Sections~\ref{sec:fp} and~\ref{sec:edf},
respectively.  As proved by Liu and Layland~\cite{Liu73}, single
processor preemptive FP and EDF have the following worst-case scenario
for the job releases
\begin{equation}
  \label{eq:wc_release}
  \forall j\in\mathbb{N},\quad r_{i,j} = j\,T_i,
\end{equation}
which corresponds to all tasks starting to release jobs simultaneously
and at the fastest rate.  This means that if the jobs generated by the
tasks are schedulable when released according
to~(\ref{eq:wc_release}), then they are always schedulable for any
releases fulfilling~(\ref{eq:min_inter}). From now on, we are then
assuming the worst-case scenario of~(\ref{eq:wc_release}).

\section{Fixed Priority}
\label{sec:fp}

When tasks are scheduled by Fixed Priority (FP), we assume they are
indexed by decreasing priority, that is $\tau_\ell$ has priority higher
than $\tau_i$ if and only if $\ell<i$.

For the purpose of optimizing over task execution time, it is
convenient to borrow the exact schedulability condition, as formulated
by Lehoczky et al.~\cite{Leh89}. 

\begin{theorem}[from~\cite{Leh89}]
  A periodic task set $\mathcal{T}$ is schedulable under Fixed
  Priority \textbf{if and only if}
  \begin{equation}
    \label{eq:iffCondCi}
    \forall i=1,\ldots,n \quad
    \exists t\in[0,D_i] \quad
    C_i+\sum_{\ell=1}^{i-1} \ceilfrac{t}{T_\ell}C_\ell \leq t.
  \end{equation}
\end{theorem}
By using the more compact vector notation, the
Eq.~(\ref{eq:iffCondCi}) can be rewritten as
\begin{equation}
  \label{eq:compactIffCspace}
  \forall i=1,\ldots,n \quad
  \exists t\in[0,D_i] \quad
  \bk_i(t)\cdot \mathbf{C} \leq t
\end{equation}
where
\[
  \bk_i(t)=\Bigg(\overbrace{\ceilfrac{t}{T_1}, \ceilfrac{t}{T_2},
    \ldots, \ceilfrac{t}{T_{i-1}}}^{\text{from $0$ to $i-1$}},1,\overbrace{0,\ldots,0}^{\text{from $i+1$ to $n$}}\Bigg).
\]

Testing if Eq.~(\ref{eq:compactIffCspace}) is true for any $t$ in the
dense interval $[0,D_i]$ is not practically feasible. In fact with
elementary considerations Lehoczky suggested the equivalent
\begin{equation}
  \label{eq:iff_RM_Leho}
  \bigwedge_{i=1}^n\ 
  \bigvee_{t\in\mathcal{S}_i} \ 
  \bk_i(t)\cdot \mathbf{C} \leq t
\end{equation}
with $\mathcal{S}_i$ being the discrete and finite set defined by
\begin{equation}
  \label{eq:Leho_points}
  \mathcal{S}_i = \{j\,T_\ell: 1\leq \ell<i,\ j\,T_\ell\leq D_i\}\cup\{D_i\}.
\end{equation}
The set $\mathcal{S}_i$ contains all the release instants $r_{j,\ell}$
of any task $\tau_\ell$ with priority higher than $\tau_i$, with
$r_{j,\ell}\leq D_i$, plus the deadline $D_i$ of the task $\tau_i$
itself.

In Equation~(\ref{eq:iff_RM_Leho}) we expressed the same condition
of~(\ref{eq:compactIffCspace}) through the logical AND/OR operator
instead of the propositional operators $\forall/\exists$. This makes
more clear that the space of FP-schedulable execution times is the
intersection of unions of the half-spaces
$\{\bk_i(t)\cdot \bC \leq t\}$ in space of execution times
$\bC\in\mathbb{R}^n$.

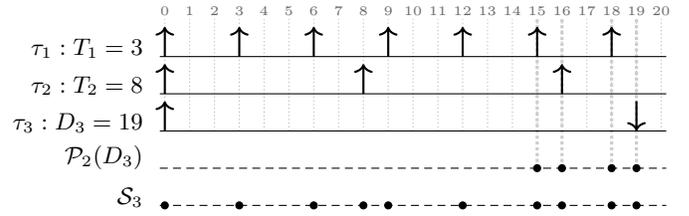
\begin{figure}[tb]
  \centering
  \begin{center}
    \begin{tikzpicture}[scale=0.33]
      \def\vert{1.5} 
      \draw foreach[count=~]\l in{0,1,...,20}
      {(\l,\vert) node[anchor=south,yshift=-2pt, black!60] {\tiny$\l$}};      
      \draw[densely dotted, black!30] (0,\vert) grid[ystep=0] +(20,{-3*\vert});

      \def\lev{-3*\vert}
      \draw[densely dashed] (-0.2,\lev) -- (20.2,\lev);
      \draw (-0.5,\lev) node[anchor=south east,yshift=-4pt]
      {\small $\mathcal{P}_2(D_3)$};
      \foreach \a in {15,16,18,19} {
        \draw[densely dotted, very thick,black!20] (\a,\vert) -- (\a,\lev);
        \draw[fill] (\a,\lev) circle [radius=4pt];
      }

      \def\lev{-4*\vert}
      \draw[densely dashed] (-0.2,\lev) -- (20.2,\lev);
      \draw (-0.5,\lev) node[anchor=south east,yshift=-4pt]
      {\small $\mathcal{S}_3$};
      \foreach \a in {0,3,6,8,9,12,15,16,18,19} {
        \draw[fill] (\a,\lev) circle [radius=4pt];
      }

      \def\lev{0}  
      \def\per{3}  
      \def\job{6}  
      \foreach \j in {0,1,...,\job}{
        \draw[->,thick] ({\j*\per},\lev) -- +(0, 0.8*\vert);
      }
      \draw (-0.2,\lev) -- (20.2,\lev);
      \draw (-0.5,\lev) node[anchor=south east,yshift=-4pt] {\small $\tau_1:T_1=\per$};
 
      \def\lev{-\vert}
      \def\per{8}  
      \def\job{2}  
      \foreach \j in {0,1,...,\job}{
        \draw[->,thick] ({\j*\per},\lev) -- +(0, 0.8*\vert);
      }
      \draw (-0.2,\lev) -- (20.2,\lev);
      \draw (-0.5,\lev) node[anchor=south east,yshift=-4pt] {\small $\tau_2:T_2=\per$};

      \def\lev{-2*\vert}  
      \def\per{100}  
      \def\dl{19}   
      \def\j{0}   
      \draw[->,thick] ({\j*\per},\lev) -- +(0, 0.8*\vert);
      \draw[->,thick] ({\j*\per+\dl},{\lev+0.8*\vert}) -- +(0, -0.8*\vert);
      \draw (-0.2,\lev) -- (20.2,\lev);
      \draw (-0.5,\lev) node[anchor=south east,yshift=-4pt]
      {\small $\tau_3:D_3=\dl$};
    \end{tikzpicture}
  \end{center}
  \caption{An example of the schedulability points $\mathcal{S}_3$ and
    $\mathcal{P}_2(D_3)$ for a set of $3$ tasks with $T_1=3$, $T_2=8$,
    and $D_3=19$. Notice that the set of points does not depend on the
    execution times.}
  \label{fig:sched_FP}
\end{figure}

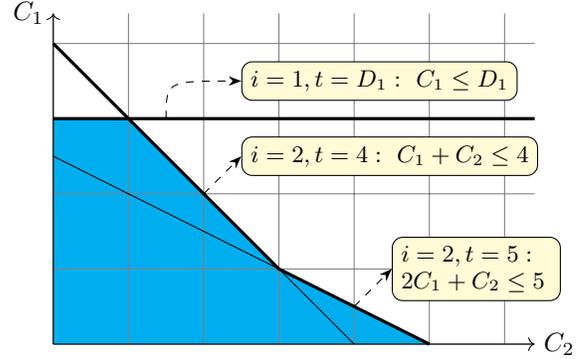
\begin{figure}[tb]
  \centering
  \begin{tikzpicture}[scale=1]
    \def\limCone{4.4}
    \def\limCtwo{6.4}

    \fill[cyan] (0,0) -- (0,3) -- (1,3) -- (3,1) -- (5,0) -- cycle;
    \draw[gray] (0,0) grid (\limCtwo,\limCone);
    \draw[->] (0,0) -- (\limCtwo,0) node[anchor=west] {$C_2$};
    \draw[->] (0,0) -- (0,\limCone) node[anchor=east]{$C_1$};

    \draw[very thick] (0,3) -- +(\limCtwo,0);
    \linOR{5}2141

    \node[draw, rounded corners, fill=yellow!20, anchor=west]
    (label) at (2.5,3.5) {\small$i=1,t=D_1:\ C_1\leq D_1$};
    \draw[-{Stealth},dashed] (1.5,3) .. controls (1.5,3.5) .. (label.west);

    \node[draw, rounded corners, fill=yellow!20, anchor=west]
    (label) at (2.5,2.5) {\small$i=2,t=4:\ C_1+C_2\leq 4$};
    \draw[-{Stealth},dashed] (2,2) -- (label.west);

    \node[draw, rounded corners, fill=yellow!20, anchor=west]
    (label) at (4.5,1) {\small\parbox{2cm}{$i=2,t=5:$\\$2C_1+C_2\leq 5$}};
    \draw[-{Stealth},dashed] (4,0.5) -- (label.west);
    
    \end{tikzpicture}
    \caption{Region of RM schedulable execution times. We draw $C_2$
      along the horizontal axis and $C_1$ along the vertical axis. In this
      example, we assume $n=2$ tasks and parameters: $T_1=4$, $D_1=3$,
      and $D_2=5$ and any $T_2\geq
      D_2$. From~(\ref{eq:iff_RM_schedP}), when $i=1$ we have
      $\mathcal{P}_0(D_1)=\{D_1\}=\{3\}$ which gives $C_1\leq 3$ as
      the only (and trivial) necessary and sufficient constraint to
      guarantee the schedulability of $\tau_1$, the highest priority
      task.  When $i=2$ the schedulability points are
      $\mathcal{P}_2(D_2)=\{5,4\}$, which yield the constraints
      $2C_1+C_2\leq 5$ and $C_1+C_2\leq 4$, respectively, both
      represented by thin lines. We need to make union among these
      constraints, as required by the logical OR
      of~(\ref{eq:iff_RM_schedP}), thus getting the two oblique thick
      segments. Since the overall schedulability region is given by the
      intersection of the single-task regions, we find that the
      RM-schedulable execution times are the ones represented in the
      cyan area.}
  \label{fig:workRM}
\end{figure}

The challenge in a direct exploitation of~(\ref{eq:iff_RM_Leho}) for
the optimization over the execution times is that the cardinality of
the set $\mathcal{S}_i$ of~(\ref{eq:Leho_points}) may grow
significantly as the periods of high priority tasks gets smaller and
smaller w.r.t.\ the deadline $D_i$. This issue was then
addressed~\cite{Man98,Bin04b}. It was demonstrated that \textbf{if the
  priorities are Rate Monotonic}, then a task set is schedulable by
RM, if and only if
\begin{equation}
  \label{eq:iff_RM_schedP}
  \bigwedge_{i=1}^n\ 
  \bigvee_{t\in\mathcal{P}_{i-1}(D_i)} \ 
  \bk_i(t)\cdot \mathbf{C} \leq t
\end{equation}
with $\mathcal{P}_i(t)$ generically defined by
\begin{equation}
  \left\{
    \begin{array}{l}
      \mathcal{P}_0(t)=\{t\}\\
      \mathcal{P}_i(t)=
      \mathcal{P}_{i-1}\left(\floorfrac{t}{T_i}T_i\right)
      \cup\mathcal{P}_{i-1}(t).
    \end{array}
  \right.
  \label{eq:definitionP}
\end{equation}
Figure~\ref{fig:sched_FP} shows an example of computation of the two
sets.  In the example with $i=3$, it can be observed that the points
in $\mathcal{P}_{2}(D_3)$ are $4$. Instead, the number of points in
Lehoczky's $\mathcal{S}_3$ are $10$.  In general, the number of
necessary and sufficient points that need to be tested
through~(\ref{eq:definitionP}) is constant with a given number of
tasks, whereas the number of points from~(\ref{eq:Leho_points}) may
grow arbitrarily large with the task set
parameters. Figure~\ref{fig:workRM} illustrates the space of RM
schedulable execution times for an example with $2$ tasks.

\subsection{Open problems}
\label{sec:openFP}

In this section we sketch some problems which, to best of our
knowledge, are open.

\paragraph*{Non-DM priorities}
The reduction of schedulability points of Eq.~(\ref{eq:definitionP})
can be made only when priorities are DM/RM~\cite{Man98,Bin04b}. The
proof does exploit the fact that higher priority tasks have a smaller
period than the one under analysis. Is a construction similar to the
one of~(\ref{eq:definitionP}) applicable to generic non-DM/RM
priorities? In some preliminary experiments~\cite{2009Gang}, it was
shown a counter-example of a non-DM tasks' set with a schedulability
point \textbf{not in} the set of~(\ref{eq:definitionP}). However, this
investigation was not continued any further and the counter-example is
lost.

\paragraph*{Tightness of the points}

The set of points determined by (\ref{eq:definitionP}) is certainly
smaller than the original Lehoczky's set of~(\ref{eq:Leho_points}). It
remains an open question if the reduced set of points can be further
reduced:
\begin{itemize}
\item without exploiting information on the execution times, and
\item keeping the set as necessary and sufficient condition.
\end{itemize}

\paragraph*{Arbitrary deadlines}

The direct extension of (\ref{eq:iff_RM_schedP}) to the arbitrary deadline case would be
\begin{equation}
  \label{eq:iff_RM_schedP_arb}
  \bigwedge_{i=1}^n\ 
  \bigwedge_{j=0}^{\mathsf{lastbusy}(i)}\ 
  \bigvee_{t\in\mathcal{P}_{i-1}(j\,T_i+D_i)} \ 
  \bk_i(t,j)\cdot \mathbf{C} \leq t
\end{equation}
with $\bk_i(t,j)$ defined by
\[
  \bk_i(t,j)=\Bigg(\overbrace{\ceilfrac{t}{T_1}, \ceilfrac{t}{T_2},
    \ldots, \ceilfrac{t}{T_{i-1}}}^{\text{from $0$ to $i-1$}},j+1,\overbrace{0,\ldots,0}^{\text{from $i+1$ to $n$}}\Bigg).
\]
properly extended to account the job $j$ of $\tau_i$. Also, in
Eq.~(\ref{eq:iff_RM_schedP_arb}), $\mathsf{lastbusy}(i)$ denotes the
index of the last $\tau_i$ job in the level-$i$ busy
interval~\cite{Leh90}. The formulation of~(\ref{eq:iff_RM_schedP_arb}), however,
poses a few challenges with no answer:
\begin{itemize}
\item is there any redundancy among the many schedulability points in
  $\mathcal{P}_{i-1}(j\,T_i+D_i)$ as $j$ spans from $0$ to
  $\mathsf{lastbusy}(i)$?
\item since the execution times $\bC$ are unknown, how long is the
  level-$i$ busy interval? In the special case with $\sum_iU_i = 1$
  and no hyperperiod $H$ (irrational periods), which implies that
  every instant is level-$n$ busy and then $\mathsf{lastbusy}(n)\to\infty$, how can we test the schedulability
  of $\tau_n$?
\end{itemize}

\section{Earliest Deadline First}
\label{sec:edf}

In this section, we illustrate some results following a similar
investigation for the EDF scheduling policy. Also, we remark that in
this section, we relax the constrained deadline case and allow
deadlines to be arbitrary (possibly larger than the period of the
corresponding task). Also we denote by $H$ the \emph{hyper-period},
which is the least common multiple among all task periods
$\{T_1,\ldots,T_n\}$. Observe that in our initial hypothesis we
assumed all parameters to be \textbf{real-valued}. Hence, we assume
that the hyperperiod $H$ exists (as it normally happens in reality)
and postpone the curious case of a non-existent hyperperiod $H$ to
Section~\ref{sec:openEDF} for a related open problem.

If the $n$ tasks in $\mathcal{T}$ are scheduled by preemptive EDF, the
following condition is necessary and sufficient to ensure that no job
deadline is missed.
\begin{theorem}[Corollary 1 in~\cite{Bar90a}]
  \label{th:exact_test}
  The task set $\mathcal{T}$ is scheduled by preemptive EDF \textbf{if
    and only if}
  \begin{equation}
    \sum_{i=1}^n U_i \leq 1,
    \label{eq:tot_uleq1}
  \end{equation}
  and
  \begin{multline}
    \forall t\in\mathbb{N}, 0\leq t\leq H+\max_i\{D_i\},\\
    \underbrace{\sum_{i=1}^n\max\left\{0,\floorfrac{t-D_i}{T_i}+1\right\}C_i}_{\mathsf{dbf}(t)}\leq t
    \label{eq:dbf_cond}
  \end{multline}
\end{theorem}

The LHS of Eq.~(\ref{eq:dbf_cond}) is a very frequently used function
in real-time EDF scheduling, and it is called \emph{demand bound
  function} $\mathsf{dbf}(t)$.
Since $\mathsf{dbf}(t)$ is piecewise constant, the inequality of
Eq.~(\ref{eq:dbf_cond}) needs to be checked only at the instants $t$
when the $\mathsf{dbf}(t)$ changes value, which are all the absolute
deadlines of any job. Hence, the exact condition of
Theorem~\ref{th:exact_test} can be simplified as stated in next
corollary.

\begin{corollary}
  \label{cor:exact_dlSet}
  The task set $\mathcal{T}$ is scheduled by preemptive EDF \textbf{if
    and only if}
  \begin{equation}
    \forall t\in\mathcal{D}, \quad \bh(t) \cdot \bC \leq 1
    \label{eq:dbf_cond_dlSet}
  \end{equation}
  with
  \begin{equation}
    \label{eq:def_dlSet}
    \mathcal{D} = \big\{d_{i,j}=j\,T_i+D_i : 
    d_{i,j}\leq H+\max_i\{D_i\}\big\}\cup\{0\}
  \end{equation}
  and $\bh(t)=[h_1(t),\ldots,h_n(t)]$, 
  \begin{equation}
    h_i(t) =
    \begin{cases}
      \frac 1{T_i}& \text{if $t=0$}\\
      \frac 1t\max\left\{0,\floorfrac{t-D_i}{T_i}+1\right\}
      &\text{otherwise}
    \end{cases}
    \label{eq:def_eta}
  \end{equation}
\end{corollary}

\begin{figure*}[t]
  \centering
  \includegraphics[width=\textwidth]{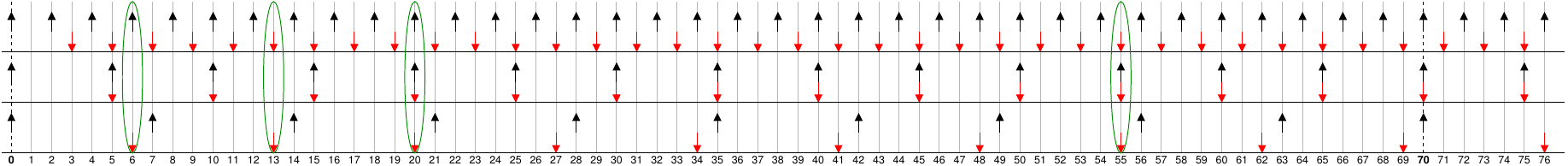}
  \caption{The tight set of necessary and sufficient constraints for
    EDF. In this example, the parameters are $T_1=2, D_2=3$,
    $T_2=5, D_2=5$, and $T_3=7, D_3=6$. Job releases are represented
    by upward black arrows. Job deadlines are represented by downward
    red arrows. We represent the deadlines until
    $H+\max\{D_i\}=70+6=76$, as required by~(\ref{eq:def_dlSet}). The
    number of total constraints to be checked is $49$ corresponding to
    $48$ deadlines plus the utilization constraint
    of~(\ref{eq:tot_uleq1}). The reduced number of constraints,
    however, is only $5$: $4$ deadlines (circled in green) plus the
    utilization constraint.}
  \label{fig:example_Dmin}
\end{figure*}
In the definition of $\mathcal{D}$ of Eq.~(\ref{eq:def_dlSet}), we use
the fictitious ``deadline  at $0$'' to encode the utilization constraint
of~(\ref{eq:tot_uleq1}) through the special definition of
$h_i(0)=\frac 1{T_i}$.

\begin{figure}[tb]
  \centering
  \begin{tikzpicture}[scale=1]
    \def\limCone{5.2}
    \def\limCtwo{7.2}
    \draw[->] (0,0) -- (\limCtwo,0) node[anchor=north] {$C_2$};
    \draw[->] (0,0) -- (0,\limCone) node[anchor=east]{$C_1$};

    \fill[cyan] (0,0) -- (0,3) -- (1,3) -- (5,0) -- cycle;

    \draw[gray] (0,0) grid (\limCtwo,\limCone);
    
    \draw (0,3) -- +(\limCtwo,0);
    \linear 115
    \linear 217

    \linear 32{11}
    \linear 43{15}  
    \linear 53{19}
    \linear 54{20}

    \draw[very thick] (0,0) -- (0,3) -- (1,3) -- (5,0) -- cycle;

    \node[draw, rounded corners, fill=yellow!20, anchor=west]
    (label) at (3,3.5) {\small$t=3:\ C_1\leq 3$};
    \draw[-{Stealth},dashed] (0.5,3) .. controls (0.5,3.5) .. (label.west);

    \node[draw, rounded corners, fill=yellow!20, anchor=west]
    (label) at (4,2.5) {\small$t=15:\ 4C_1+3C_2\leq 15$};
    \draw[-{Stealth},dashed] (3,1.5) -- (label.west);
    
    \end{tikzpicture}
    \caption{Region of EDF schedulable execution times. As in
      Figure~\ref{fig:workRM}, we draw $C_2$ along the $x$ axis and
      $C_1$ along the $y$ axis. Using the same example of
      Figure~\ref{fig:workRM}, we assume $2$ tasks with parameters:
      $T_1=4$, $D_1=3$, and $T_2=D_2=5$. In this case, the set
      $\mathcal{D}$ from~(\ref{eq:def_dlSet}) of all deadlines to be
      considered is $\mathcal{D}=\{3, 5, 7, 10, 11, 15, 19, 0\}$. We
      remind that the ``deadline $0$'' represents the utilization
      constraint of~(\ref{eq:tot_uleq1}). All constraints are
      represented by a thin line, whereas the boundary of their
      intersection is represented by a thicker line. It can be
      observed that a large majority of the constraints does not
      contribute to determine the boundary of EDF-schedulable
      execution times. In this example, only the $2$ deadlines at $3$
      and at $15$ are needed to characterize the exact region.}
  \label{fig:workEDF}
\end{figure}
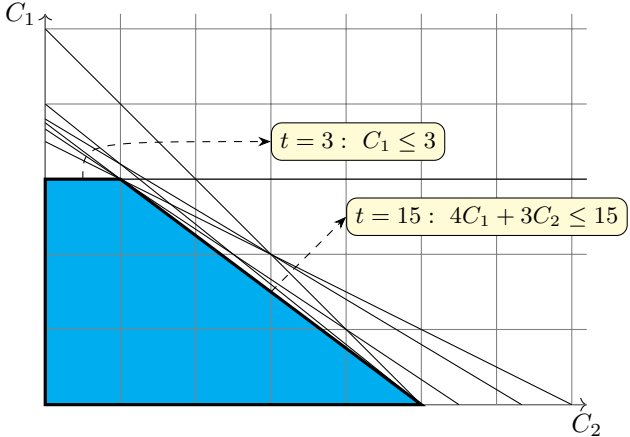

The necessary and sufficient schedulability condition
of~(\ref{eq:dbf_cond_dlSet}) already gives some information on the
geometry of the EDF-schedulable computation times. In fact, the space
of EDF-schedulable execution times is \textbf{convex} as it is the
intersection between linear halfspaces yielded
by~(\ref{eq:dbf_cond_dlSet}). Figure~\ref{fig:workEDF} illustrates the
EDF-schedulable execution times of the same simple example of
Figure~\ref{fig:workRM}.

As Figure~\ref{fig:workEDF} seems to suggest, there may be some
deadlines in $\mathcal{D}$ which are not necessary to be checked, as
implied by other constraints. Most of the methods to reduce the number
of deadlines to be checked~\cite{Bar90a,Rip96,Zha09}, and then the
complexity of the test exploit the values of the execution times with
the general trend that ``the smaller the utilization
$\sum_iU_i$ of the whole set of tasks, the fewer deadlines are
necessary''. However, the question of whether some deadlines can be
eliminated from $\mathcal{D}$ without compromising the schedulability
and without exploiting the execution times received little attention.

In a recent work~\cite{bini2019cutting}, it was proposed a method that
exploits the convex hull of vectors to determine the smallest subset
of deadlines $\mathcal{D}_\mathsf{min}\subseteq\mathcal{D}$ such that
\begin{equation}
  \label{eq:constraints_tight}
  \forall t\in\mathcal{D}_\mathsf{min}, \quad \bh(t) \cdot \bC \leq 1.
\end{equation}
On the one hand this method shows that the number of necessary and
sufficient linear constraints for EDF schedulability is orders of
magnitude less than the full set of~(\ref{eq:dbf_cond_dlSet}). See
Figure~\ref{fig:example_Dmin} for an example. On the other hand it
works as a ``black box'' algorithm producing the minimal set of
constraint, providing no insight on why these few deadlines only
matters. Also, the complexity of computing the convex hull of a large
set of vectors is very high.

\begin{figure}[tb]
  \centering
  \begin{tikzpicture}
    \node[anchor=south west,inner sep=0] (image) at (0,0)
    {\includegraphics[scale=0.32]{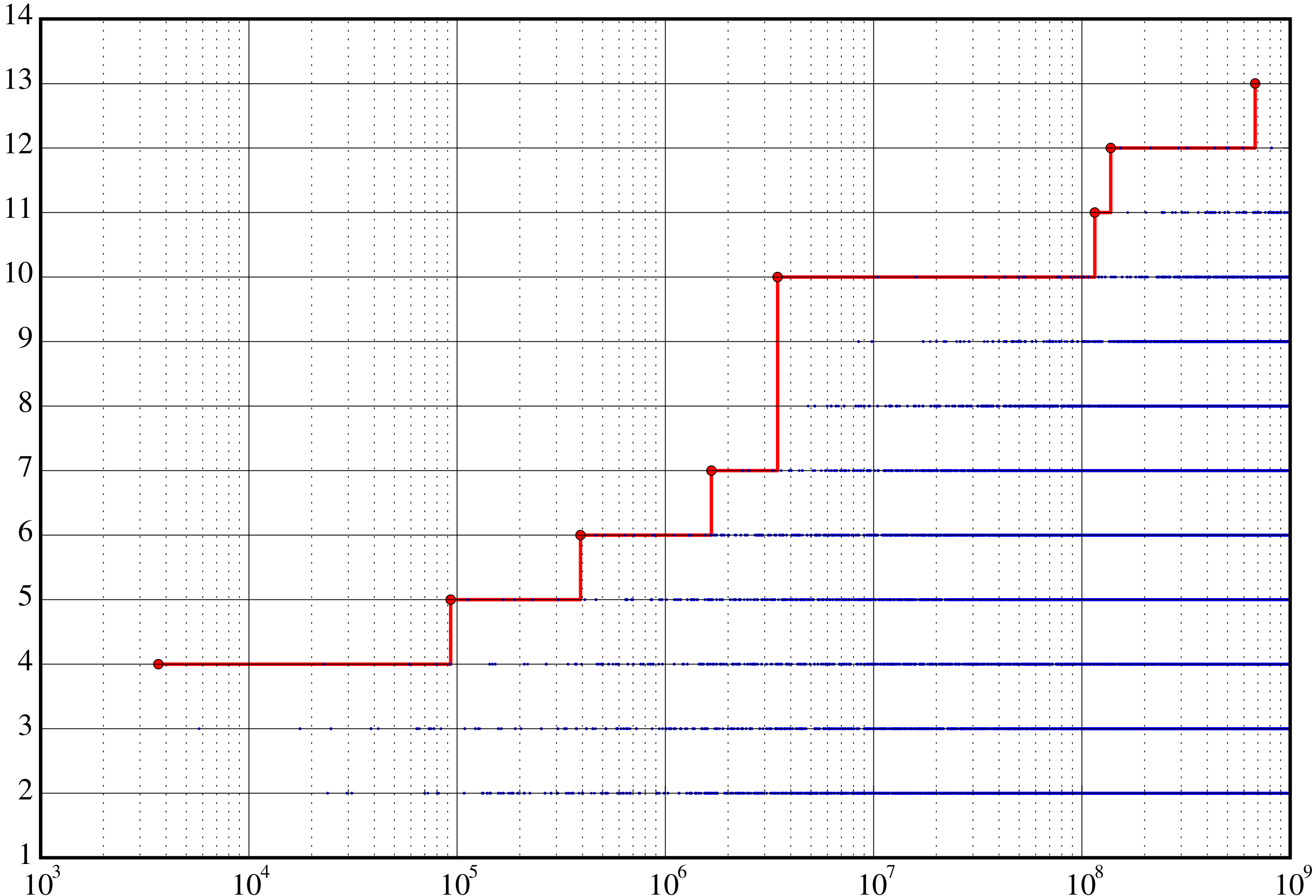}};
    \begin{scope}[x={(image.south east)},y={(image.north west)}] 
      \node[anchor=north] at (0.5,0) {Hyperperiod $H$};
      \node[anchor=south east,rotate=90] at (0,1) {Cardinality of $\mathcal{D}_\mathsf{min}$};
    \end{scope}
  \end{tikzpicture}
  \caption{In this experiment, randomly generated integer task period
    and deadlines for $n=2$ tasks. For each experiment, we are
    plotting a dot at the corresponding hyperperiod $H$ along the
    horizontal axis in log scale, and the number of necessary and
    sufficient constraints along the vertical axis. At the right, also
    the density of the minimal number of constraints over the sample
    space. We also plot the upper envelope of the points as such an
    envelope represents the hardest instances to be tested.}
  \label{fig:minEDFpoints_n2}
\end{figure}

\begin{figure}[tb]
  \centering
  \begin{tikzpicture}
    \node[anchor=south west,inner sep=0] (image) at (0,0)
    {\includegraphics[scale=0.32]{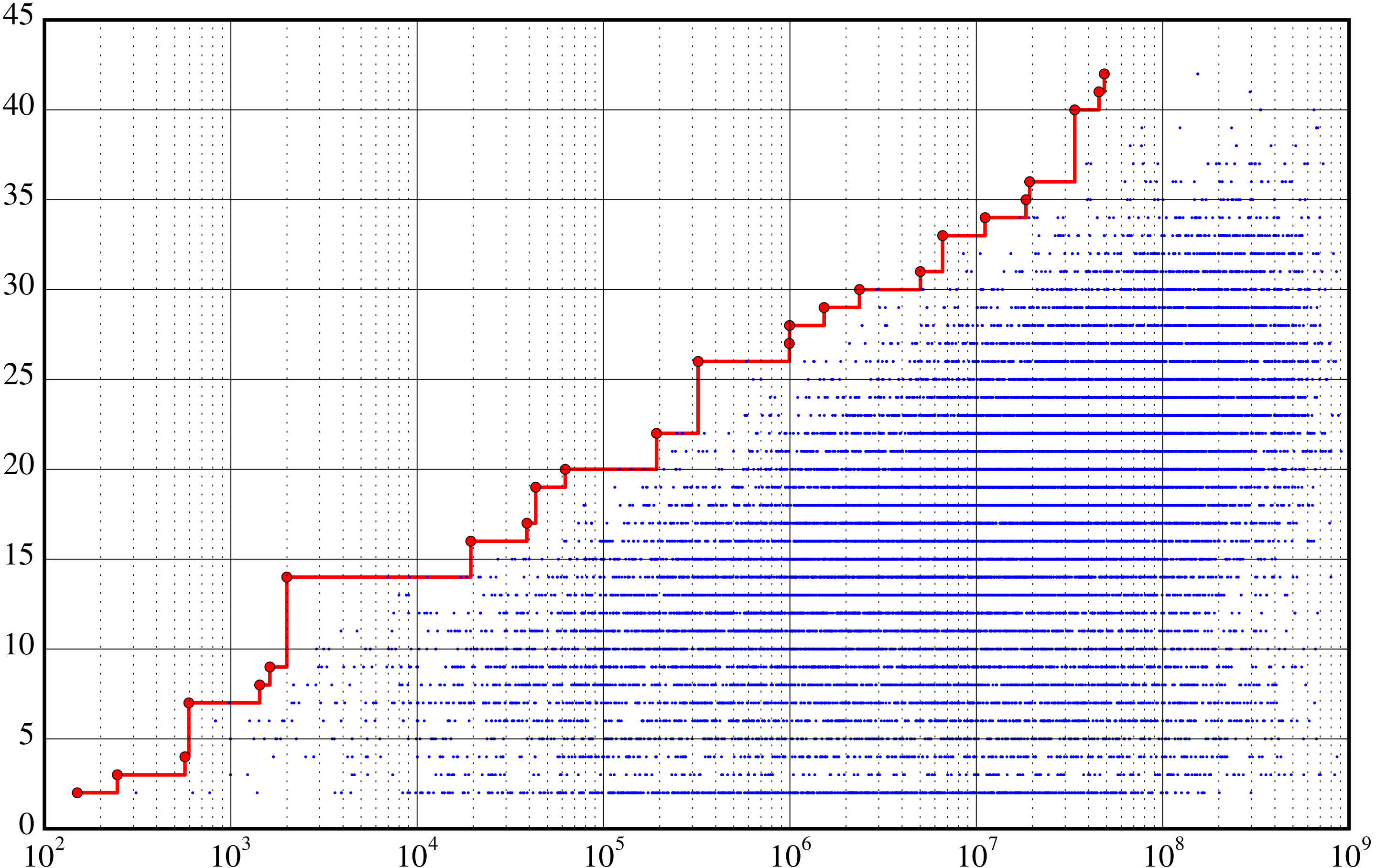}};
    \begin{scope}[x={(image.south east)},y={(image.north west)}] 
      \node[anchor=north] at (0.5,0) {Hyperperiod $H$};
      \node[anchor=south east,rotate=90] at (0,1) {Cardinality of $\mathcal{D}_\mathsf{min}$};
    \end{scope}
  \end{tikzpicture}
  \caption{Same experiment of Figure~\ref{fig:minEDFpoints_n2} with $n=3$ tasks.}
  \label{fig:minEDFpoints_n3}
\end{figure}

\begin{figure}[tb]
  \centering
  \begin{tikzpicture}
    \node[anchor=south west,inner sep=0] (image) at (0,0)
    {\includegraphics[scale=0.32]{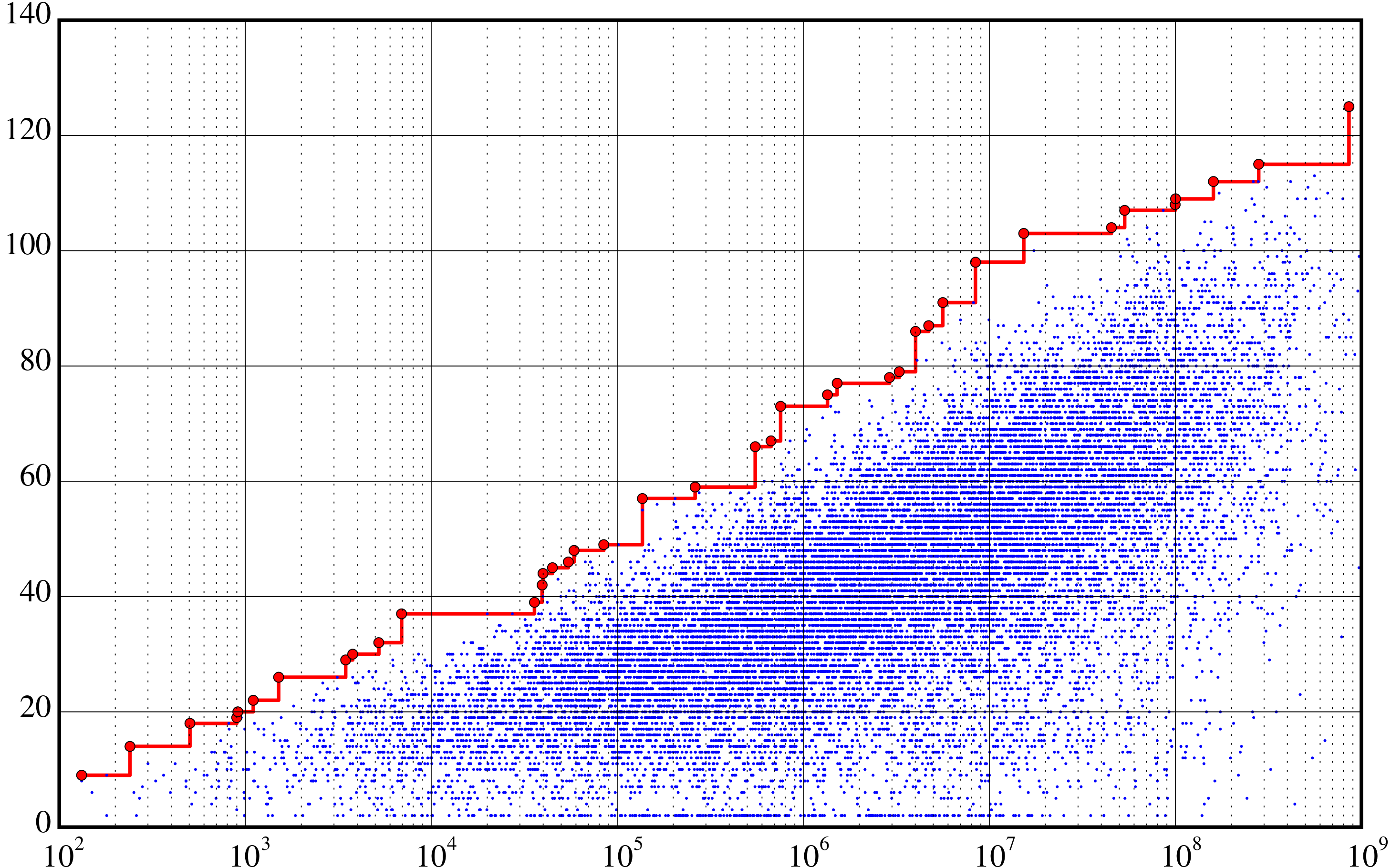}};
    \begin{scope}[x={(image.south east)},y={(image.north west)}] 
      \node[anchor=north] at (0.5,0) {Hyperperiod $H$};
      \node[anchor=south east,rotate=90] at (0,1) {Cardinality of $\mathcal{D}_\mathsf{min}$};
    \end{scope}
  \end{tikzpicture}
  \caption{Same experiment of Figure~\ref{fig:minEDFpoints_n2} with
    $n=4$ tasks.}
  \label{fig:minEDFpoints_n4}
\end{figure}

In Figures~\ref{fig:minEDFpoints_n2},~\ref{fig:minEDFpoints_n3}
and~\ref{fig:minEDFpoints_n4} we experimentally investigate the
dependency of the number of necessary and sufficient constraints on
the hyperperiod $H$. In red, we draw the upper envelope, which then
represents the hardest cases to be analyzed.  Despite the linear
growth of $|\mathcal{D}|$ with $H$ as implied by~(\ref{eq:def_dlSet}),
the number of necessary and sufficient constraints of the hardest
cases grows with $\log H$. If this experimental evidence becomes a
confirmed fact, it may allow the existence of an exact EDF test which
is polynomial in the task periods~\cite{2019Ekberg}, deferring then the
harder complexity to the number $n$ of tasks only. This seems not to
contradict any existing result.

\subsection{Open problems}
\label{sec:openEDF}

\paragraph*{Non-existent hyperperiod $H$}
If the task periods are real-valued, indeed their least common
multiple, the hyperperiod $H$, may not exist. In this case, the number
of deadlines to be tested is infinite. Also, the experiments of
Figures~\ref{fig:minEDFpoints_n2}, \ref{fig:minEDFpoints_n3}, and
\ref{fig:minEDFpoints_n4} indicates that as $H$ grows, the number of
necessary and sufficient constraints grows (logarithmically) with
$H$. Some existing EDF sufficient tests~\cite{Dev03,Cha06,Bin09f} do
not require the existence of an hyperperiod. Perhaps, some of them
become exact as $H$ tends to infinity? Anyhow, how to test EDF
schedulability with no hyperperiod $H$ is unknown.

\paragraph*{Other approaches to minimal set of constraints}
The employment of the convex hull to determine the tight set of
points~\cite{bini2019cutting} is indeed very complex. Is there any
logic behind the points selected by the convex hull? Why are deadlines
at $6$, $13$, $20$, and $55$ in the example of
Figure~\ref{fig:example_Dmin} so special? If such a logic is found,
then we could go straight to these constraints, with no complex
machinery as the convex hull. Also, it may be possible that such an
algorithm is polynomial in the periods as it grows with $\log(H)$.

\section{Related works}
\label{sec:related}

The research touched by this paper covers a very broad spectrum. As
such it is surely incomplete.  Next we report our best attempt to
cover the related literature.

In Fixed Priority, the response time analysis
(RTA)~\cite{Jos86,Har87,Aud93} is indeed very widely used.  Many works
have addressed the efficiency of the RTA by proposing a later initial
instant for the iterations. In same cases, such initial instant
depends on the execution times, hence it is unsuitable for
optimization~\cite{Sjo98,Bri03}. Other works have proposed an initial
start instant for RTA, which is independent of the tasks' execution
times~\cite{lu2007period,davis2008efficient}.  Either way, the
iterative nature of RTA makes it unfit for optimization unless costly
binary search is employed~\cite{Pun97}.

If the periods have some good harmonic properties~\cite{Kuo91}, it is
possible to discard some of the points in~(\ref{eq:definitionP}) as
shown by Zeng and Di Natale~\cite{zeng2012efficient}. However, in the
general case of periods not dividing each other, it is unknown if the
same simplification is possible.

In the context of optimization of task parameters, the task model with
imprecise computation~\cite{Shi89,Chu90} was perhaps among the first
ones to allow tasks to have a variable execution time.  Reward-based
scheduling was also a very good method to decide the duration of each
individual job in EDF, assuming that the longer a job executed the
more ``reward'' is accumulated~\cite{Ayd01}.

Exploiting the set of reduced schedulability
points~\cite{Man98,Bin04b}, Bini et al.~\cite{Bin08} proposed to
perform the sensitivity analysis on the task parameters, providing a
closed-form expression for the acceptable margins for the execution
times.

The elastic task model~\cite{But02,Cha06} was also introduced as a way
to adjust the parameters (the task periods) while preserving
schedulability.

The reduction of constraints for EDF schedulability was also addressed
by George and Hermant~\cite{Lau09}. They proposed to solve an instance
of a LP problem for each absolute deadline. As shown in their paper,
however, their method is not capable to automatically cut all
unnecessary deadlines.

The investigation of the case of irrational periods received, with no
surprise, little attention from the research community. To best of our
knowledge, the only known partial result is the computation of the
task response time as supremum of the response time among all
jobs~\cite{Bin15a}.

Finally, it is worth mentioning the work by
Singh~\cite{singh2024cutting}, who proposed an interesting unification
between FP and EDF schedulability tests.

\paragraph*{Thanksgiving}
The author would like to warmly thank the reviewers for the time spent
in making a careful and detailed review of this paper.

\paragraph*{Acknowledgment}
This work is partially supported by the project ``Trustworthy
Cyber-Physical Pipelines'', funded by the MAECI Italy-Sweden
co-operation id.\ PGR02086, and the spoke ``FutureHPC and BigData'' of
the ICSC --- Centro Nazionale di Ricerca in High-Performance
Computing, Big Data and Quantum Computing funded by European Union ---
NextGenerationEU.

\balance


\end{document}